\documentstyle[12pt,aaspp4]{article}

\lefthead{Reid. Gizis}
\righthead{Low-mass Hyades binaries}
\begin {document}
\slugcomment{To appear in December 1997 A.J.}
\title{ Low-mass Binaries in the Hyades - A scarcity of brown dwarfs
\footnotemark
\footnotetext{Based on observations with NASA/ESA Hubble Space Telescope,
obtained at the Space Telescope Science Institute, which is operated by 
AURA, Inc., under NASA contract NAS5-26555}}

\author {I. Neill Reid and John E. Gizis}
\affil {Palomar Observatory, 105-24, California Institute of Technology, 
Pasadena, CA 91125, e-mail: inr@astro.caltech.edu, jeg@astro.caltech.edu}

\begin {abstract}

We have obtained {\it HST} Planetary Camera observations of a total of
fifty-three low-mass ($M  < 0.3 M_\odot$) members 
of the Hyades cluster. Nine of these stars are resolved as binaries, with
separations between 0.1 and 3.1 arcseconds, while a further three are probably
equal-mass systems at smaller separations. Allowing for observational selection
effects, this corresponds to an observed binary fraction of $11.3 \pm 4.6$\% 
for systems with separations in the range
14 to 825 a.u., consistent with observations of solar neighbourhood M-dwarfs.
The mass-ratio distribution is only consistent marginally with the secondary
stars being drawn from a $\Psi(M) \propto M^{-1}$ mass function, with three equal-mass
systems amongst the six binaries with observed separations in the r\'egime
where our observations extend below the hydrogen-burning limit. Considering
the entire sample, the absence of any brown dwarf companions amongst our
sample makes it unlikely that the mass function of stellar/brown dwarf companions
rises as steeply as M$^{-1}$. If the Hyades has an age of $\sim 600$ Myrs, our
results are consistent only at the 2$\sigma$ level with a flat (M$^0$) mass function.

\end {abstract}

\keywords {binaries: general; stars: low mass; stars: luminosity function}

\section {Introduction }

It is by now well established that approximately half of the stellar systems
in the Galactic disk consist of at least two stars. This high frequency of
multiple stars suggests that binary formation is an integral part of the overall
star-formation process, rather than separate, post-formation event. Thus, an empirical
determination of parameters such as the overall binary frequency, the mass-ratio
distribution and the variation in those quantities as a function of the mass 
of the primary star, can constrain formation mechanisms. Moreover, binary
star surveys represent one of most effective methods of searching for objects
of substellar mass, and hence setting limits on the likely numbers of brown
dwarfs.

Most statistical studies of binaries have concentrated on field stars in the 
immediate solar vicinity. In particular, Duquennoy \& Mayor (1991) conducted a detailed spectroscopic
survey of 164 late-F and  G-dwarfs drawn from the Nearby Star Catalogue (\cite{gl69}; \cite{gj79}).
Combining the spectroscopic binaries detected from those observations with data 
from the literature on binaries at wider separations, they estimate a binary
frequency of at least 60\% amongst solar-type stars. In contrast, surveys of lower-mass
stars (\cite{mb89}, \cite{hm90}, \cite{shk96}) indicate a lower overall binary
frequency of $\sim 35\%$. Moreover, \cite{rg97} have argued that the mass ratio
distribution amongst M-dwarfs differs significantly from that amongst the higher-mass
G-dwarfs, with a clear bias towards equal-mass systems amongst the low-mass stars.
This conclusion rests upon observations of a total of barely 75 M-dwarfs, however.

Open clusters provide an alternative avenue for studying binary statistics. One
has the advantage of a sample of comparable size to that provided by the
nearest stars, but comprising stars of known age and abundance, formed in
the same environment. One is, however, faced with the potential drawback 
that dynamical evolution may have influenced either (or both) the relative
binary fraction and the distribution of orbital parameters. The latter issue can be
addressed by combining observations of stars in clusters spanning a range
of ages.

Amongst the nearer open clusters, the Hyades, at an average distance of 48 parsecs,
is clearly a prime target. The proximity of the cluster and the significant
space motion relative to the Sun lead to substantial proper motions, and surveys
by \cite{vb52}, \cite{p70}, \cite{lhm81}, \cite{sch91}, \cite{r92} and \cite{bhj94}
have provided a well-defined list of cluster members to V=19.5 or M$_V \sim 16$. 
Radial velocity surveys for spectroscopic binaries have been undertaken, primarily by
\cite{g88}, while high-resolution imaging surveys for companions
at wider separations have been carried out at optical (\cite{m93}) and infrared
(\cite{p97}) wavelengths. These surveys, however, are restricted to the
brighter stars in the cluster, scarcely extending fainter than K-type dwarfs.
Thus, while the results provide a valuable dataset for comparison with the
\cite{dm91} analysis, there is little overlap with the local surveys of low-mass stars.

Over the last three years, we have been carrying out a survey aimed specifically
at identifying binary companions at moderate to wide separations amongst the
lowest-mass members of the Hyades cluster. We have been using the Planetary Camera 
on the Hubble Space Telescope to obtain high spatial-resolution images which
are capable of detecting secondary companions at separations of 15 a.u. or more, 
and with masses at or below the hydrogen-burning limit. Preliminary results from
this survey were reported in \cite{gr95} (hereinafter paper I). 
Observations have since been made of a further 39 candidate Hyads, 
more than doubling the sample size. Section 2
describes those observations; in section 3, we calculate the binary frequency and
compare our result against observations of higher-mass Hyades stars, and data
for stars in the Solar Neighbourhood; section 4 discusses the implications of
our lack of detection of any candidate brown dwarf companions; section 5
presents our conclusions.

\section {Observations }

\subsection {HST Planetary Camera data}

All of the stars in the present sample are selected from the region of
the Hyades covered by the \cite{r92} survey. All have VRI photometry (Reid, 1993)
and absolute magnitudes of M$_V \ge +11.9$. Table 1 lists 
photometry and (for Hyades stars) distance estimates for the thirty-nine 
stars added to the sample since the observations reported in paper I.
All save one of these stars are drawn from the \cite{r92} proper-motion
survey: the exception is B 804, from \cite{bhj94}, which is 
the lowest-luminosity Hyades member identified to date. The distances are
derived from the observed proper motion, assuming that the stars are
Hyades members. The \cite{r92} motions are absolute, calibrated using a
galaxian reference frame, and there is good agreement between the 
mean cluster motion derived from that study and Schwann's (1991) meridian circle
results. On that basis, we have adopted the Schwann determination of the
convergent point and the \cite{det84} cluster radial velocity measurement 
for these calculations; our previous analysis was based on the \cite{gunn} 
convergent point. Schwann's analysis leads to a distance modulus of 3.40
magnitudes (47.9 parsecs) to the cluster centre, and provides a distance-scale 
zeropoint accurate to $\sim 3\%$. 
Distances derived for individual stars using this method are independent of the photometry, 
and the $\sim 0\farcs005$ uncertainties in the proper motions of these faint stars
correspond to a $\sim5\%$ uncertainty in the distance. 

Each star was centred in the Planetary Camera (PC) on HST, and observed using the 
F850LP filter with exposure times of from 3 to 70 seconds. The latter were
specified to leave the target star unsaturated, allowing a direct measurement
of the relative brightness of any potential companions detected. Two images were taken
of most stars from our Cycle 4 allocation and of all stars observed during Cycle 6,
allowing unambiguous elimination of cosmic-ray events.  The PC
has a plate-scale of 0.046 arcsec pixel$^{-1}$, while the diffraction limit of 
HST at 8500 \AA is 0.073 arcseconds. As described in \cite{rg97}, we have
used both the DAOFIND routines and Lucy-Richardson deconvolution techniques to 
search for binary companions. These methods are capable of detecting an
equal-luminosity companion at a separation of 0.09 arcseconds, and companions
fainter by $\Delta I$ of 1, 3 and 5 magnitudes at separations of 0.14, 0.23 and 0.31
arcseconds, respectively. Three stars in the current sample (RHy 126, 309 and 391)
appear to have significant elongation in our PC observations, suggesting that 
they are probably equal-mass binaries with angular separations of $\sim 0.1$
arcseconds. Table 1 lists our estimates of $\rho$ and $\Delta I$ for those
three stars, where we have used DAOPHOT point-spread function fitting to
deconvolve the HST images. The dates of each binary-star observation are
listed in Table 1, and the images are available in the HST archive.

Four of the stars in the present sample are identified as background
field stars, rather than Hyades cluster members. As Table 1 show, all 
four stars have formal membership probabilities of less than 20\% and, 
as Hyades members, lie towards the lower edge of the colour-magnitude relation (Reid, 1993).
(As emphasised above, the distances, and hence absolute magnitudes, are derived
from the observed proper motion under the assumption that the transverse velocity 
matches that of the cluster.) Two stars, RHy 80 and 271, 
lack respectively the chromospheric and coronal activity which characterises
cluster members\footnote {We note that RHy 225, with a
membership probability of 89\%, and included as a cluster member in paper I,
also lacks H$\alpha$ emission, and is likely to be a non-member. Cluster
membership probability is based on comparing the stellar proper motion against
a model for the expected cluster distribution and the proper motions of the
underlying field star. Observationally, we deal with individual stars, rather
than distributions, and there is nothing to prevent a field star having exactly 
the motion expected for a cluster member. Thus it is essential that proper motion-selected
stars be regarded as only candidate cluster members, regardless of the formal membership
probability, until confirmed by independent photometric, chromospheric and radial
velocity criteria.}. The
other two stars, RHy 77 and 110, are multiple systems, and are rejected as cluster members 
based on their location in the H-R diagram after allowing for the contribution
to the ground-based photometry made by the secondary star (Figure 1a). RHy 77 is a near
equal-mass binary, separation 0.46 arcseconds, while RHy 110 is a triple system, with 
components of I=15.32, 16.21 and 17.00 mag ($\rho$(AB)=0\farcs69, $\Delta$(BC)=0\farcs17).

The widest binary listed in Table 1 has a separation of 1.12 arcseconds. None of
these binaries were resolved in ground-based observations, and with HST observations at
only one epoch, we cannot identify these stars conclusively as bound companions.
However, the observed surface density of stars with 14 $<$ I $<$ 17 magnitude is only
$\sim 1600$ deg$^{-2}$ in the Hyades fields surveyed by \cite{r92}, 
implying an average separation of 3 arcminutes between
stars, and a probability of $\sim 6 \times 10^{-4}$ of finding two stars at
1.2 arcseconds separation. All of the resolved binaries listed in Table 1 are
therefore very likely to be bound systems. 

Possible companions at wider separations are accessible to ground-based
CCD and photographic observations, notably POSS I and POSS II plate
material. All of the potential wide binary companions detected in the
PC observations are visible on at least the POSS II, and usually the
POSS I, R-band plate material. None have either colours that are sufficiently
red to be consistent with late-type M-dwarfs, or proper motions consistent
with cluster membership. 

\subsection {The colour-magnitude diagram}

Figure 1a plots the (M$_V$, (V$-$I)) colour-magnitude diagram for M-dwarfs that
are identified as Hyades members based on their proper motions (Reid, 1993, 
appendix\footnote{In addition to the late-type M-dwarfs cited in section 2.1, 
five stars listed as possible members by \cite{r93} have
since been identified as field stars: vA 106 (G 7-150), G 7-116, RHy 61 (LP 474-1156), 
RHy 131 and RHy 158 (LP 415-1198).}). 
The fainter stars with HST observations are identified separately
in the figure. All of the binaries except RHY 240AB (the two lowest
luminosity binary components, see paper I) have joint photometry, and
most lie significantly above the main sequence. We have used the HST
observations to estimate the relative contribution of each binary
component to the ground-based V \& I data, and Figure 1b plots the results.
We assume equal-mass components for each of the possible binaries listed
in Table 1. Lacking high-resolution V-band data, we have plotted the
primary stars at the joint photometry colours - in most cases, the
components are of nearly equal-mass and the colours can be expected
to be little-affected. We have estimated (V$-$I) colours for the
secondary stars (excepting RHY 240AB) based on the mean colour-magnitude 
relation.

The reduced scatter in Figure 1b is partly a
result of these approximations. However, it is also clear that the 
main-sequence is noticeably narrower once due allowance is made for
the composite nature of the HST binaries. There are a number of other 
obvious candidate binaries, notably RHy 390 at M$_V$=15.03, (V$-$I)=3.83, but
also amongst earlier-type M-dwarfs. We are
currently obtaining high-resolution spectroscopic observations, using HIRES
on Keck I, with the aim of identifying close stellar binaries amongst the
lower-luminosity stars.

Until recently, the usual representation of the (M$_V$, (V$-$I)) lower main-sequence was 
a linear relation. However, \cite{ldh94} first pointed out that the
Hyades colour-magnitude relation exhibited significant non-linearities 
at fainter magnitudes. 
\cite{gr96} demonstrated that there was a substantial discontinuity in the
absolute magnitude/TiO bandstrength relation at M$_V \sim +12$; and \cite{rg97}
have shown that a break occurs at similar luminosities in the field-star (M$_V$, (V$-$I))
lower main-sequence. \cite{cl97} have argued that this feature is not due to
atmospheric effects (i.e., structure in the effective temperature/colour relation),
but rather reflects a change in the internal structure. Figure 1b plots the 
three-segment colour-magnitude relation derived by \cite{rg97} from V- and I-band 
photometry of Solar Neighbourhood stars. The higher-metallicity Hyades stars are
redder than local field stars of the same absolute magnitude but, 
simply offsetting the nearby-star relation by 0.17 magnitudes
in (V$-$I) to fit the Hyades main-sequence at M$_V < 11$ leads to a reasonable
match to the general morphology of the data at fainter magnitudes. \cite{cl97}
have suggested that the change in slope at M$_V \sim +12$ reflects a change in the
mass-radius relation, perhaps as a result of the stars becoming fully convective.
Whether this hypothesis is confirmed or not, cluster colour-magnitude 
diagrams offer the possibility of tracing the position of this break as a function
of abundance and age. Reliable identification of such features, however, demands both 
accurate photometry and an adequate multiple-star census, as demonstrated by Figures
1a and 1b.
   
\section {The Binary Fraction }

Combining the present observations with the results from paper I, we have
identified at least nine, and perhaps as many as twelve, binaries amongst
our total sample of fifty-three Hyades M-dwarfs. This corresponds to a
binary fraction, f$_{ovbs}$ of between 17 and 23\%, with formal uncertainties of $\pm 7\%$
through counting statistics. These estimates are consistent with the results
derived in our preliminary analysis (Gizis \& Reid, 1995). Clearly, this represents 
a lower limit to the the overall binary fraction, f$_{tot}$, since only companions 
with projected separations of at least 0.1 arcseconds can be resolved 
unambiguously by the PC. Indeed, as
described in the previous section, the limiting magnitude for detecting a companion
star varies with separation due to the underlying point-spread function 
of the primary star. Thus, transforming the observed binary fraction to an
estimate of the total binary fraction requires a model for the semi-major axis
distribution so that one can allow for for unresolved systems at small angular separation. 
In paper I we combined an estimate of the detection efficiency
as a function of radius with two representations of the semi-major axis
distribution (the \cite{dm91} G-dwarf log-normal period distribution, and 
Fischer \& Marcy's (1992) results for field M-dwarfs) to derive the
appropriate correction factor. Applying that factor to the value of f$_{obs}$ derived
in paper I leads to an inferred f$_{tot}$
of $27 \pm 16$\%, with the substantial uncertainties reflecting the
small sample size. Our present data allow a modest improvement in the
statistical accuracy.

We adopt a slightly different strategy in the analysis in the current paper.
Theoretical models calculated by \cite{bhsl} predict that the hydrogen-burning
limit in the Hyades corresponds to M$_{bol} \sim 12.8$. The 
predicted effective temperature is $\sim 2700$ K, comparable to that of GJ 1111,
for which \cite{leg96} derive an I-band bolometric correction of 0.3 magnitudes. 
This implies that the hydrogen-burning limit corresponds to M$_I \sim 13.1$ magnitudes,
4 magnitudes fainter than the brightest stars in our sample. On the basis of our analysis 
of the HST images, we would expect to detect a companion with $\Delta I = 4$ mag at
a separation of $\rho \sim 0.28$ arcseconds, or 13.5 a.u. at a distance of 48 parsecs. 
Hence, our observations are capable of detecting all stellar companions with separations
of between 13.5 and 825 a.u., where the upper limit is set by the field of view of
the PC. These constraints define a complete sample of six binaries
(RHy 49, 119, 221, 240, 246 and 377 - Table 2), or a binary fraction of 11.3$\pm 4.6$\%. 

We can compare this statistic against the
distribution of $\rho$ amongst the binaries in the 8-parsec sample defined by
\cite{rg97}. In doing so, we take 825 a.u., corresponding to the semi-diameter of the
PC field of view, as the upper limit to $\rho$ in the Hyades distribution. Gliese 644/VB 8 is
the only local binary that lies outwith this limit. Limiting the calculation 
to the 78 systems where the brightest star has M$_v \ge 8.0$, 11 of the 24 known
binary (or multiple) stars have components separated by at least 13.5 a.u. Thus, 
f$_{obs}$ is $14.1 \pm 4.2\%$, consistent with our Hyades observations.
It therefore follows that if the cluster binaries have a semi-major axis distribution
that is comparable with the field M-dwarfs, then the implied f$_{tot}$ is
$25 \pm 10$\%. \cite{mcin} have pointed out that there is an apparent deficit of
wide binaries ($\rho >$ 100 a.u.) amongst the Hyades stars when compared with the
field-star sample. This might reflect disruption of such loosely-bound systems over
the lifetime of the cluster. 

Our results for the late-type Hyades stars can also be matched against the binary
statistics for higher-mass cluster members. Patince et al (1997) have completed infrared speckle 
observations of 162 main-sequence Hyades stars with $K < 8.5$ (M$_K < 5, M_V < 7$),
observations capable of detecting binary companions with $\rho > 0.1$ arcseconds
and $\Delta K < 4$ magnitudes. Eleven of the 162 stars (or 6.8\% of the sample) 
have resolved companions within the range of angular separation matching the HST
PC observations. Since the K-band mass-luminosity relation can be approximated
by M$_K \propto 0.15 \times$ log(mass), a magnitude difference of $\Delta K = 4$ corresponds 
to a mass ratio of $\sim$0.25. Thus, unlike our HST observations, in most cases the
speckle data are incapable of detecting companions with masses close to
the H-burning limit. One can allow for this incompleteness by adopting a 
mass function for the stellar companions. If that mass function matches the
local field-star $\Psi(M)$, then the observed binary fraction underestimates the
true binarity {\sl for this range of separations} by a factor of 
two (see \cite{p97} for full details). In that
case, the f$_{obs}$ amongst the higher-mass cluster-members for 
companions with $14 < \rho < 825$ a.u. is $13.5 \pm 4$\% - consistent with our 
observations of the late-type M-dwarfs in the cluster. With the relatively
small number of stars in either sample, however, the statistical significance of 
this result is not strong, 

\section {The mass-ratio distribution and brown dwarfs}

In their analysis of the later-type binaries amongst stars within 8-parsecs
of the Sun, \cite{rg97} argued that there was evidence for a bias towards
equal-mass systems; that is, the hypothesis that both the primary and secondary
star are drawn at random from the same mass function was not consistent with the
observed distribution of mass-ratio, q=$M_2 \over M_1$. 
We can test this result using our Hyades observations.

Table 2 lists mass estimates for each star in the resolved systems. These are 
based on the bolometric magnitudes, derived from M$_I$ using the \cite{leg96} 
bolometric corrections. For stars fainter
than M$_{bol}$ = 10, we have used the 600-Myr X-models calculated by \cite{bhsl} 
to estimate masses, while masses for more luminous stars are estimated from an
empirical relation calibrated by the \cite{hm93} nearby-star speckle data.
For these calculations, we again limit the sample to the six binaries which fall 
in the r\'egime where our data are capable of reaching the hydrogen-burning limit.
The mass function for the local stars is well-represented by a power-law. 
$\Psi(M) \propto M^{-1}$ (Reid \& Gizis, 1997). Dynamical effects are likely to have
modified the overall Hyades mass function, with preferential evaporation of
lower-mass stars. However, while wide binaries may have been stripped, 
it is unlikely that the cluster dynamical
evolution influences the companion-star mass distribution. 

We have used Monte-Carlo techniques to model the 
expected mass-ratio distribution amongst binary stars, generating 
companions (mass $M_C$) to each star (mass M$_*$)
in the current sample, with $\Psi(M) \propto M^{-1}$ and $M_* \ge M_C \ge 0.08 M_\odot$.
Figure 2 compares the predicted mass-ratio distribution, F(q), plotted as
a cumulative distribution, against the observed distribution. With only six binaries
in the latter sample, the statistical significance of the comparison is not 
overwhelming. However, three of the six stars are effectively equal-mass
systems, and a Kolmogorov-Smirnoff test indicates that there is only a 10\% chance
of the observed distribution being drawn from the distribution predicted if the
companion stars are selected from a $\Psi(M) \propto M^{-1}$ mass function. Thus,
these data are consistent with the hypothesis that there is a bias toward
equal-mass systems amongst M-dwarf binaries.

The HST imaging data are also capable of detecting higher-mass brown dwarfs. 
Three stars listed in Table 2 have companions with masses in the range
0.11 to 0.13 M$_\odot$, although all are clearly above the hydrogen-burning
limit\footnote{The candidate substellar companion to RHy 281, reported
in paper I, was re-observed by HST and is not a common
proper-motion companion of the Hyades star.}. With a detection limit 
of $\Delta I$ = 5 magnitudes for $\rho >0.31$ arcseconds, our deepest
observations reach magnitude limits fainter than I=19.5, corresponding to
M$_I \sim 16$, or M$_{bol} \sim 15.5$. In the \cite{bhsl} models
this luminosity corresponds to a brown dwarf of mass 0.05 M$_\odot$. 

Since our observations extend beyond the hydrogen-burning limit, we
can use these data to constrain the form of $\Psi_S (M)$, the secondary
star mass function. If we adopt a functional form for $\Psi_S (M)$, we
can calculate the expected relative number density of very low-mass (VLM)
stars and substellar objects contributed by each star in our sample:
that is, if $\Psi_S (M) \propto M^{-1}$ and the appropriate limiting
magnitude corresponds to M$_I = 17$, or $ M \sim 0.05 M_\odot$, then

$$ {{N(0.25 \ge M/M_\odot < 0.1)} \over {N(0.1 \ge M/M_\odot > 0.05)}} \quad = \quad
{{\int_{0.1}^{0.25} \Psi_S (M) dM } \over{{\int_{0.05}^{0.1} \Psi_S (M) dM }}} \quad \sim
\quad 1.3 $$

That is, with a mass function increasing as $\Psi (M) \propto M^{-1}$ towards
lower masses, one predicts almost equal numbers of companions with masses
between 0.05 and 0.1 M$_\odot$, and between 0.1 and 0.25 M$_\odot$. Given that
each HST observation has a different limiting magnitude, we can determine the 
corresponding mass limit, and derive the appropriate ratio for that star. Summing the
contribution from every star in the sample gives the expected relative numbers
of stars and brown dwarfs detected in the present survey.

Since our observations are in the I-band, we need to determine the
appropriate bolometric corrections to estimate the mass-limit corresponding to
the limiting magnitude of each PC observation. A 0.055 M$_\odot$ Hyades brown
dwarf is predicted to have a temperature of $\sim 1700$ K (Burrows et al, 1993) and
bolometric corrections remain poorly defined for objects of such low temperatures.
However, GD 165B is estimated to have a temperature of $\sim 1800 \pm 100$ K
(\cite{ts96}), and \cite{ti93} have determined a bolometric correction of
BC$_I$=1.75 magnitudes for that star. Lacking other empirical results, we
adopt a bolometric correction of BC$_I$=0.3 magnitudes at M$_I$=13 mag and assume
a linear increase by 0.4 magnitudes per magnitude at fainter magnitudes. 

We can used the mass-luminosity relation defined by the \cite{bhsl} models to
transform the M$_{bol}$ detection-limit to a mass limit. However, these
models are valid only if the Hyades has an age of 600 Myrs.
While some recent analyses (e.g. \cite{tor96}) favour ages close to this value, 
other colour-magnitude diagram studies derive ages older by $\sim 300$ Myrs 
(Friel, 1995). Clearly, the older the cluster, the lower the luminosity achieved
by brown dwarfs of a given mass. To take this uncertainty into account, we
have also estimated mass-limits using the relation

$$ { L \over L_\odot} \quad = \quad 3.4 \times 10^{-6} \quad ({ t \over {10^9}})^{-1.3} \quad
( {M \over {20 M_J}} )^{2.64} $$

where t is the age in years, and M the mass in Jovian units. This relation, 
from \cite{bl93}, is based on brown dwarf cooling models, and 
is expected to be valid for brown dwarfs older than $\sim 10^8$ years. There
are relatively small differences between the Burrows {\sl et al.} X-models and
this mass-luminosity relation for an age of 600 Myrs. 

Our calculations are therefore based on three brown dwarf mass-luminosity
relations: the Burrows {\sl et al.} X-models, and the \cite{bl93} relation for ages of 600 Myrs
and 1 Gyr, taking the last as an upper limit to the cluster age.
Table 3 lists the results, where $\alpha$ is the power-law exponent adopted for the
mass function. We have divided the 0.25 to 0.05 M$_\odot$ mass range at 0.10
and 0.15 M$_\odot$, and compare the observed and predicted numbers above (*) and
below (VLM) the divide. The complete sample of six binaries defined above provide the
empirical reference. While that reference sample is small in number, it
is clear that, even for an age of 1 Gyr, a companion-star mass function 
$\Psi_S (M) \propto M^{-1} $ predicts almost $\sim 2 \sigma$ too many objects
with masses below 0.15 M$_\odot$. If the 
Hyades age is close to 600 Myrs, then even a flat mass function ($\propto M^0$) is
inconsistent with the observations at the same level. These results, therefore,
favour the hypothesis that both VLM dwarfs and brown dwarfs are infrequent 
companions to lower-mass M-dwarfs.

\section {Conclusions }

We have completed our {\it HST} Planetary Camera search for lower-luminosity
companions of a sample of fifty-three low-mass ($M  < 0.3 M_\odot$) members 
of the Hyades cluster. Each observation is capable of detecting
companions with masses at the hydrogen-burning limit where the projected angular 
separation is at least 0.28 arcseconds. Setting the outer radial limit at
the circle circumscribed by the PC image, this corresponds to linear separations
of from $\sim 14$ to 825 a.u. at the average distance of the Hyades cluster.
In the case of the lowest luminosity
stars, we expect to be able to detect brown dwarfs with masses as low
as 0.05 M$_\odot$. However, within the annulus where our observations
are complete, we identify only six stellar companions. The binary
fraction is comparable with that observed amongst local M-dwarfs in the
same range of separations and, as with the local stars, there is a
statistically-significant preference for equal-mass systems and a
scarcity of both very low-mass M-dwarf and brown dwarf companions. 

\acknowledgments {This research was funded by HST grant GO-5353.01-93A
and GO-06344.01-95A. 
This work is based partly on photographic plates obtained at the Palomar
Observatory 48-inch Oschin Telescope for the Second Palomar
Observatory Sky Survey which was funded by the Eastman Kodak
Company, the National Geographic Society, the Samuel Oschin
Foundation, the Alfred Sloan Foundation, the National Science
Foundation grants AST84-08225, AST87-19465, AST90-23115 and
AST93-18984,  and the National Aeronautics and Space Administration 
grants NGL 05002140 and NAGW 1710.  The first Palomar 
Observatory Sky Survey was funded by the  
National Geographic Society, and the Oschin Schmidt Telescope is  
operated by the California Institute of Technology and Palomar Observatory.
}

\clearpage

\begin{deluxetable}{lrrrcrrrrrrr}
\tablecaption{HST observations of candidate low-mass Hyads}

\tablewidth{0pt}
\tablenum{1}
\tablehead{
\colhead{Name} & \colhead {V} &\colhead {(V-I)} &
\colhead{M$_I$} & \colhead {comments} &
\colhead{M$_I$(A)} & \colhead{$\rho$} & \colhead{$\theta$} &
\colhead{M$_I$(B)} }
\startdata
 & & & & Binaries &\nl
  RHy  49    &  15.89 &   3.04 &   9.37 &   r=49.7 pc.&10.00 & 0\farcs36 &103& 11.65 \nl 
  RHy 119    &  17.29 &   3.54 &  10.52 &   44.3&  11.23 & 0.88 &187& 11.32\nl 
  RHy 221    &  15.88 &   3.10 &   9.34 &   48.8&  10.19 & 0.31 &80& 10.22\nl
  RHy 244    &  15.64 &   2.97 &   9.00 &   54.2& 9.33 & 0.14 &162& 10.45 \nl 
  RHy 346    &  16.03 &   3.18 &   9.56 &   45.5& 10.12 & 0.48 &250& 10.55\nl 
 & & & & Binaries? &\nl
  RHy 126    &  17.33 &   3.38 &  10.07 &   59.7 pc& 11.25 & $\sim$0.06 &171&11.25 \nl 
  RHy 309    &  16.75 &   3.24 &   9.39 &   66.7& 10.15 & $\sim$0.05 &57&10.15\nl 
  RHy 391    &  16.42 &   3.35 &   9.63 &   48.8& 10.35 & $\sim$0.08 &131& 10.35\nl 
 & & & & Single stars &\nl
  RHy  23    &  16.18 &   3.01 &   9.62 &   51.3 pc& \nl
  RHy  46    &  16.68 &   3.12 &   9.86 &   55.0& \nl
  RHy  64    &  16.33 &   3.23 &  10.32 &   36.0& \nl
  RHy 101    &  17.11 &   3.46 &  10.64 &   40.0& \nl
  RHy 115    &  16.50 &   3.03 &   9.61 &   59.2& \nl
  RHy 128    &  17.63 &   3.34 &  10.77 &   50.6& \nl
  RHy 143    &  16.23 &   3.02 &   9.71 &   50.1& \nl
  RHy 162    &  16.54 &   3.21 &  10.32 &   40.0& \nl
  RHy 163    &  15.44 &   3.00 &   9.38 &   40.9& \nl
  RHy 182    &  15.89 &   3.06 &   9.37 &   49.2& \nl
  RHy 199    &  16.70 &   3.22 &  10.10 &   47.4& \nl
  RHy 200    &  16.07 &   2.89 &   9.57 &   52.7& \nl
  RHy 206    &  16.63 &   3.09 &  10.06 &   49.7& \nl
  RHy 219    &  16.37 &   2.87 &  10.06 &   48.8& \nl
  RHy 230    &  18.25 &   3.42 &  11.20 &   53.2& \nl
  RHy 231    &  15.89 &   3.11 &   9.28 &   50.1& \nl
  RHy 242    &  15.91 &   2.92 &   9.53 &   49.2& \nl
  RHy 260    &  16.40 &   3.07 &  10.35 &   39.4& \nl
  RHy 297    &  17.04 &   3.18 &  10.14 &   55.5& \nl
  RHy 298    &  15.90 &   2.93 &   9.62 &   46.8& \nl
  RHy 312    &  15.21 &   2.86 &   9.17 &   43.3& \nl
  RHy 331    &  16.43 &   2.99 &  10.21 &   44.3& \nl
  RHy 369    &  16.14 &   2.92 &   9.57 &   53.7& \nl
  RHy 376    &  15.80 &   2.90 &   9.52 &   47.4& \nl
  RHy 399    &  15.09 &   2.92 &   9.16 &   40.0& \nl
  RHy 402    &  17.77 &   3.25 &  10.59 &   61.1& \nl
   B  804    &  19.27 &   3.87 &  11.96 &   48.8& \nl
 & & & & Non-members &\nl
RHy  77 & 17.37 & 2.81 & & P=19\%$^1$, binary &\nl
RHy 80 &  17.94 & 3.17 & & P=3\%, no H$\alpha$ &\nl
RHy 110 & 16.78 & 3.03 & & P=8\%, triple &\nl
RHy 271 & 17.86 & 3.09 & & P=3\%, no X-ray &\nl
\enddata
\tablenotetext{1}{The formal probability, based on proper
motion alone, that the star is a member of the Hyades cluster.}
\tablecomments{ HST observations of binary stars were made on the following
dates: RHy 49 - 25:01:96; RHy 119 - 15/02/96; RHy 126 - 19/11/95; 
RHy 221 - 23/02/96; RHy 244 - 13/09/95; RHy 309 - 03/02/96; RHy 346 - 
25/10/96; RHy 391 - 04/02/96.}
\end{deluxetable}

\begin{deluxetable}{rrrccccc}
\tablecaption{Binary star masses}

\tablewidth{0pt}
\tablenum{2}
\tablehead{
\colhead{RHy} & 
\colhead{M$_I$(A)} & \colhead{M$_I$(B)} & \colhead{$\Delta$ (a.u.)} &
\colhead{$M_A$ (M$_\odot$)} & \colhead{$M_B$ (M$_\odot$)} &
\colhead{ q} }
\startdata
49 & 10.20 & 10.33 & 17.9 & 0.17 & 0.17 & 0.98 &\nl
119 & 11.04 & 11.32 & 39.0 & 0.13 & 0.11 & 0.88 &\nl
202 & 9.88 & 11.66 & 12.3 & 0.21 & 0.10 & 0.51 &\nl
221 & 10.08 & 10.11 & 15.1 & 0.19 & 0.18 & 0.99 &\nl
240 & 11.53 & 11.62 & 152.7 & 0.11 & 0.11 & 0.98 &\nl
244 & 9.18 & 10.45 & 7.6  & 0.25 & 0.16 & 0.65 &\nl
346 & 10.06 & 10.55 & 21.8 & 0.19 & 0.16 & 0.86 &\nl
371 & 9.26 & 10.59 & 6.8 & 0.25 & 0.21 & 0.70 &\nl
377 & 9.68 & 10.62 & 66.4 & 0.22 & 0.15 & 0.70 &\nl
\enddata
\end{deluxetable}

\begin{deluxetable}{lcccccccc}
\tablecaption{Predicted numbers of VLM stars and brown dwarfs}

\tablewidth{0pt}
\tablenum{3}
\tablehead{
\colhead{} & \colhead {$\alpha$} & \colhead { N (M$_*^1$)} 
& \colhead { N (M$_{VLM}^1)$} &\colhead {  }& \colhead { N (M$_*^2$)} 
& \colhead { N (M$_{VLM}^2)$} }
\startdata
 &\nl
Observations & & 6 & 0 & &4 & 2&\nl
 & \nl
Analytic & 0 & 6 & 1.1 & &4 & 3.1 &\nl
600 Myr & -0.5 & 6 & 1.6 && 4 & 4.3 &\nl
    & -1 & 6 & 2.2 & &4 & 5.9 &\nl
    & -1.5 & 6 & 3.1 && 4 & 8.2 &\nl
 & \nl
Analytic & 0 & 6 & 0.4 && 4 & 2.3 &\nl
1 Gyr & -0.5 & 6 & 0.5 && 4 & 3.0 &\nl
  &   -1 & 6 & 0.7 && 4 & 3.9 &\nl
    & -1.5 & 6 & 1.0 && 4 & 5.1 &\nl
 & \nl
Burrows et al & 0 & 6 & 1.8 && 4 & 4.4 &\nl
X-files &-0.5 & 6 & 1.9 && 4 & 4.6 &\nl
  & -1 & 6 & 2.7 && 4 & 6.4 &\nl
    & -1.5 & 6 & 3.7 && 4 & 9.0 &\nl
\enddata
\tablecomments{ The relative numbers of companions within specific mass ranges
predicted by power-law mass functions, compared with the observed number
of companions (line 1). \newline
$\alpha$ : the power-law index of the mass function \newline
N (M$_*^1$) : number of secondaries with $0.25 \ge {M \over M_\odot} > 0.10$ \newline
N (M$_{VLM}^1$) : number of secondaries with $0.10 \ge {M \over M_\odot} > 0.05$ \newline
N (M$_*^2$) : number of secondaries with $0.25 \ge {M \over M_\odot} > 0.15$ \newline
N (M$_{VLM}^2$) : number of secondaries with $0.15 \ge {M \over M_\odot} > 0.05$ }
\end{deluxetable}

\clearpage

\pagestyle{empty}
\begin{figure}
\figurenum{1}
\plotfiddle{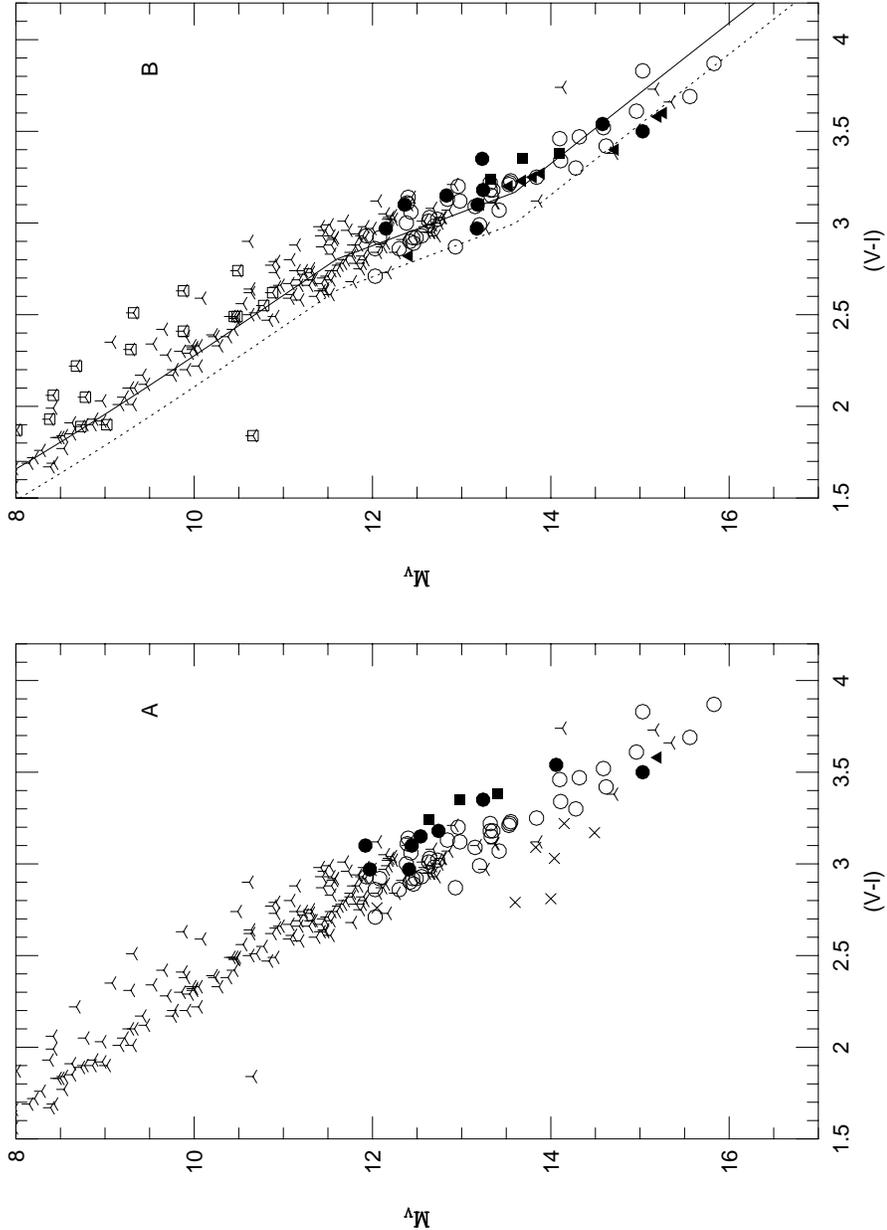}{6in}{0}{70}{70}{-240}{-30}
\caption{The Hyades colour-magnitude diagram. a) Open circles are Hyades members
with no resolved companion in the HST images; solid circles are stars with resolved
companions; solid squares are possible binaries; four-point crosses are stars 
identified as nonmembers; and three-point crosses mark data for cluster members 
with no HST observations. The solid triangle marks RHy 240B, the only secondary star
with separate ground-based photometry. The VI photometry is taken directly from Reid (1993) and 
Leggett {\sl et al.} (1994). 
b) The symbols have the same meaning as in figure 1a, save that the open squares
identify known spectroscopic binaries amongst the brighter stars, and we have used
the HST observations to estimate deconvolved magnitudes and colours for the binaries.
The primary stars are plotted as solid circles while the secondaries are plotted as
solid triangles. The dotted
line plots the three-segment relation matched to the local stars by Reid \& Gizis (1997).
The solid line plots the same relation offset by 0.17 magnitudes.}
\end{figure}

\begin{figure}
\figurenum{2}
\plotfiddle{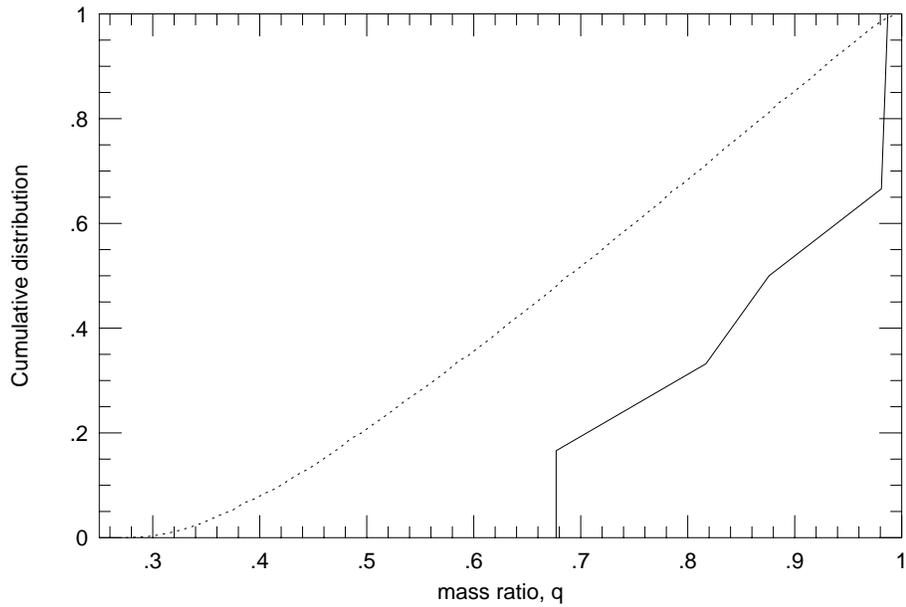}{6in}{0}{70}{70}{-230}{-30}
\figcaption { Mass-ratio distribution: the dotted line is the expected mass-ratio distribution
for the HST sample if companion stars have a mass function which varies as M$^{-1}$.
The calculations were truncated at the hydrogen-burning limit. The 
solid line is the observed mass-ratio distribution of the six binaries in our complete
sample. A Kolmogorov-Smirnoff test shows that there is only a 10\% chance of the
latter distribution being drawn from the predicted distribution. }
\end{figure}


\clearpage

 
\begin {thebibliography}{DUM}


\bibitem [Burrows {\sl et al.} (1993)] {bhsl} Burrows, A., Hubbard, W.B., Saumon, D. \& Lunine, J.I. 1993, \apj,
406, 158

\bibitem [Burrows \& Liebert (1993)] {bl93} Burrows, A. \& Liebert, J. 1993, J. Rev. mod. Phys. 65, 301

\bibitem [Bryja {\sl et al.} (1994)] {bhj94} Bryja, C., Humphreys, R.M. \& Jones, T.J. 1994, \aj, 107, 246

\bibitem [Clemens {\sl et al.} (1997)] {cl97} Clemens, J.C., Reid, I.N., Gizis, J.E. \& O'Brien, M.S. 1997, \apj. 
submitted

\bibitem[Detweiler {\sl et al.} (1984)] {det84} Detweiler, H.L., Yoss, K.M., Radick, R.R. \& Becker, S.A.,
1984, \aj, 1038

\bibitem [Duquennoy \& Mayor] {dm91} Duquennoy, A. \& Mayor, M.  1991, \aap, 248, 485

\bibitem [Fischer \& Marcy (1992)] {fm92} Fischer, D.A. \& Marcy, G.W. 1992, \apj, 396, 178

\bibitem [Friel (1995)] {fr95} Friel, E.D. 1995, \araa, 33, 381

\bibitem [Gliese, 1969] {gl69} Gliese, W. 1969, Veroff. Astr. Rechen-Instituts, Heidelberg, Nr. 22

\bibitem [Gliese \& Jahreiss, 1979] {gj79}  Gliese, W. \& Jahreiss, H. 1979, \aaps, 38, 423

\bibitem [Gizis \& Reid (1995)] {gr95} Gizis J. \& Reid, N. 1995, \aj, 110, 1248

\bibitem [Gizis \& Reid (1996)] {gr96} Gizis J.E. \& Reid, I.N. 1996, \aj, 111, 365

\bibitem [Griffin {\sl et al.} (1988)] {g88} Griffin, R.F., Gunn, J.E., Zimmerman, B.A. \& Griffin, R.E.M.
1988, \aj, 96, 172

\bibitem [Gunn {\sl et al.} (1988)] {gunn} Gunn, J.E., Griffin, R.F., Griffin, R.E.M. \& Zimmerman, B.A.
1988, \aj, 96, 198

\bibitem [Henry \& McCarthy, 1990] {hm90} Henry, T.J. \& McCarthy, D.W.  1990, \apj, 350, 334

\bibitem [Henry \& McCarthy (1993)] {hm93} Henry, T.J. \& McCarthy, D.W.  1993, \aj, 106, 773

\bibitem [Leggett {\sl et al.} (1996)] {leg96} Leggett, S.K., Allard, F., Berriman, G., Dahn, C.C. \&
Hauschildt, P.H. 1996, \apjs, 104, 117

\bibitem [Leggett {\sl et al.} (1994)] {ldh94} Leggett, S.K., Harris, H.C. \& Dahn, C.C. 1994, \aj, 108, 944

\bibitem [Lutyen {\sl et al.} (1981)] {lhm81} Luyten, W.J., Hill, G. \& Morris, S. 1981, Proper Motion Survey
with the 48-inch Schmidt Telescope, LIX, Univ. Minnesota, Minneapolis

\bibitem [Macintosh {\sl et al.} (1997)] {mcin} Macintosh, B.A., Zuckerman, B. \& Becklin, E.E. 1997, \apj, in press

\bibitem [Marcy \& Benitz, 1989] {mb89} Marcy, G.W. \& Benitz, K. 1989, \apj,  344, 441

\bibitem [Mason {\sl et al.}, 1993] {m93} Mason, B.D., McAlister, H.A., Hartkopf, W.I. \& Bagnuolo, Jr., W.G. 1993
\aj, 105, 220

\bibitem [Patience {\sl et al.}, 1997] {p97} Patience, J., Ghez, A., Reid, I.N., Matthews, K.M. \&
Weinberger, A. 1997, in preparation

\bibitem [Pels {\sl et al.} (1975)] {p70} Pels, G., Oort, J.H. \& Pels-Kluyer, H.A. 1975, \aap, 43, 423

\bibitem [Reid (1992)] {r92} Reid, I.N. 1992, \mnras,  257, 257

\bibitem [Reid (1993)] {r93} Reid, I.N. 1993, \mnras,  265, 785

\bibitem [Reid {\sl et al.} (1995)] {rhm95} Reid, I.N., Hawley, S.L. \& Mateo, M. 1995, \mnras,  272, 828

\bibitem [Reid \& Gizis (1997)] {rg97} Reid, I.N., Gizis, J.E. 1997, \aj,  113, 2246

\bibitem [Schwann (1991)] {sch91} Schwann, H. 1991, \aap, 243, 386

\bibitem [Simons {\sl et al.}, 1996] {shk96} Simons, D.A., Henry, T.J. \& Kirkpatrick, J.D. 1996, \aj, 112, 2238

\bibitem [Stobie {\sl et al.} (1987)] {s87} Stobie, R.S., Ishida, K. \& Peacock, J.A.  1989, \mnras, 238, 709.

\bibitem [Tinney {\sl et al.} (1993)] {ti93} Tinney, C.G., Mould, J.R. \& Reid, I.N. 1993, \aj, 105, 1045

\bibitem [Torres {\sl et al.} (1996)] {tor96} Torres, G., Stefanik, R.P. \& Latham, D.W. 1997, \apj, 479, 268

\bibitem [Tsuji {\sl et al.}, 1996] {ts96} Tsuji, T., Ohnaka, K., Aoki, W. \& Nakajima, T. 1996, 
\aap, 308, L29

\bibitem [van Buren (1952)] {vb52} van Bueren, H.G. 1952, BAN, 11, 385

\end{thebibliography}
\end{document}